# A Goal Question Metric Approach for Evaluating Security in a Service Oriented Architecture Context

Meryem Kassou, Laila Kjiri

Al-Qualsadi Research & Development Team, Ecole Nationale Supérieure d'Informatique et d'Analyses des Systèmes,
ENSIAS, Université Mohamed V- Souissi
BP 713, Rabat, Morocco

**Abstract**

For interactions to be possible within the Service Oriented Architecture (SOA) ecosystem, each actor must be enough confident of other actors to engage safely in the interactions. Therefore, the establishing of objective metrics tailored to the context of SOA that show security of a system and lead to enhancements is very attractive. The purpose of our paper is to present a GQM (Goal Question Metric) approach based on Standard security metrics and on SOA maturity that can be a support for organizations to assess SOA Security and to ensure the safety of their SOA based collaborations.

***Keywords:*** *SOA, Maturity Models, Security, Metrics, Assessment frameworks, Goal Question Metric approach.*

## 1. Introduction

While SOA is adopted by more organizations as an architectural style for their applications, they have to face security issues brought by related SOA context. Indeed, any SOA implementation must overcome the challenges to security of information systems, namely authentication, authorization, integrity and confidentiality in a flexible and highly distributed environment [1].

Moreover, organizations that want to do business with, or otherwise cooperate with other organizations are seeking assurance that their partners have a certain level of security in their systems. Companies are coming under increasing compliance pressures that require them to prove due diligence when protecting their data assets [2].

Security metrics are then needed to understand current state-of-security, to improve that state, and to obtain resources for improvements [3]. Many standard security assessment frameworks can support evaluating security of organizations through measuring and monitoring their security program effectiveness or security processes maturity; measuring and comparing some specific technical objects or measuring and managing the risks of operating systems [4,5,6]. But they suffer from limitations inherent either to their general purpose, ambiguity or specialized nature [7].

The purpose of our paper is to present a Goal Question Metric approach that will support the definition of tailored security metrics to assess Information System Security of Organizations that use SOA as a basis for their collaborations. We will start by presenting SOA characteristics, SOA Security challenges and SOA Maturity Models. Then, we will discuss about the use of Security Frameworks like ISO 27002 and SSE-CMM to assess security of Information Systems and how to derive security metrics. We will introduce our GQM approach to derive security metrics related to the SOA Context provided by its maturity level. Finally, we will use this approach based on a selected SOA maturity model and on Security standards, to define security goals related to access control features and to deduce related security metrics.

## 2. SOA vs. Security

### 2.1. Service Oriented Architecture Elements

SOA [8] is an architectural style that is based on reusable and loosely coupled software resources (called services) to allow the flexibility of business applications in an interoperable and open system. Figure 1 presents an SOA architectural stack; its elements are detailed hereafter [8].

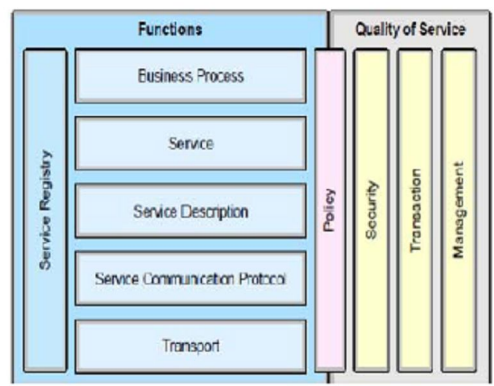

Fig.1 SOA Architecture Elements [8].

**Transport** is the mechanism used to route requests and responses to services. **Communication protocol service** is an agreed protocol used to enable communication between the service provider and customer service. **Description of service** is an agreed scheme to describe the service, its operations and its features. **Service** is the application component. **Business process** is a collection of services invoked in a particular sequence under specific rules to meet a business requirement. **Service registry** is a repository of service and data descriptions to allow the publication and service discovery and **Policy** is a set of conditions and rules used by the service provider to make the service available to consumers. **Security** is the set of rules that might be applied to the identification, authorization, and access control of service consumers invoking services. **Transaction** is the set of attributes that might be applied to a group of services to deliver a consistent result. **Management** is the set of attributes that might be applied to managing the services provided or consumed.

SOA security should be a primary consideration when establishing communications between distributed systems service based [9]. In figure 1, Security is presented as a QOS issue that is related to all the elements presented in the functions part and might be tradeoff with other quality issues.

### 2.2. SOA Security challenges and requirements

Successfully implemented SOA security has to be well-defined, well-planned, and well-implemented [9].

Like Distributed Systems, SOA security is based on several important requirements, including [10,11]:
- **Identification and Authentication:** Verifying the identity of a user, process, or device, before allowing access to resources in an information system.
- **Authorization.** The permission to use a computer resource, granted, directly or indirectly, by an application or system owner.
- **Integrity.** The property that data has not been altered in an unauthorized manner while in storage, during processing, or in transit.
- **Non-repudiation.** Both parties are able to provide legal proof to a third party that the sender did send the information, and the receiver received the identical information.
- **Confidentiality.** Preserving authorized restrictions on information access and disclosure, including means for protecting personal privacy and proprietary information
- **Auditing.** All transactions are recorded so that problems can be analyzed after the fact.
- **Privacy.** Restricting access to subscriber or relying party information in accordance with law and organization policy.

Security in SOA is more complex than traditional IT application security. Following Items should also be considered [10, 11]:
- All entities in SOA must have identities and decouple identities from applications.
- Proper security control must be applied for each service in composite applications.
- Security management across diver's environment.
- Protection of business data in transit and at rest.
- Compliance with a growing set of regulations.

### 2.3. SOA Security solutions and considerations

There are different ways to address security requirements in an SOA environment. We present below web services related security specifications and solutions, reviewed from the literature [8,9,10,11,12,13] and classified according to Architecture elements of figure1.

**At the transport level**: services are secured using the in-built security features of transport channel technologies such as HTTPS.

**At the service communication protocol level:** security at this level is ensured using SOAP message based security [14] that protects messages by encrypting and/or digitally signing the body, headers, attachments, and any combination or part thereof.

**At the service description level:** security properties are published in the interface description contract for other services to invoke upon. This contract may be implicit like the one described by the Web Service Description Language WSDL[15] or more detailed like the one provided by ebXML[16]. Other techniques can provide tools for negotiating security requirements like Semantic Web technologies [17].

**At the service level:** Service-level security includes all security mechanisms that are coupled directly with the application logic whether coded into the service component or delegated to security-specific services.

**At the Business Process Level:** Security requirements are defined by Business Rules that provide a means to express and specify high-level security constraints in the form of policy. Related work reviewed from literature focused on three points: (i) Languages to specify business process and related security constraints (WS-BPEL [18] and its security extensions), (ii) techniques to generate security implementations from abstracted security requirements (Model Driven security [19]), (iii) enriching contracts description with security semantics to enable dynamic discovery binding and negotiation of security properties.

**Security External to the SOA system:** service is loosely-connected to the security implementation through a messaging interface. An example for this is XML Firewall that is deployed at the network perimeter and enforces security policies by processing incoming and outgoing messages.

**Policy considerations:** Policy is a broad term covering not only security, but other domains such as reliability, transactions, privacy, and so on. Each domain requires a language for describing quality of service (QoS) requirements and capabilities associated with services, such as WS-SecurityPolicy [20] for security.

**Service discovery considerations:** In order to assure that the services provided are legitimate, the user should be able to authenticate the service discovery service. The service discovery also should be able to verify the authenticity of the user requesting a list of services and restrict the items seen on the list according to the authorization of the user. In addition, the service discovery must only list the services that have been verified as legitimated services.

**Management considerations:** To build secure SOA applications, the engineering process should take the security considerations into design, implementation, management and maintenance, etc. For instance [21] proposes a process for SOA Security to be used in Projects. Other management considerations are related to the monitoring, logging and audit of Security incidents.

There are other considerations to be taken:

**Assets Considerations :** Asset-level security refers to protecting any assets used by and encapsulated in the service like application data, devices and capabilities but also information describing the services, taxonomies, policy repositories, etc.

**Application front end's considerations:** it can be both an interface for a human user or another machine**.** It is unclear, how information provided to a frontend, is used in the following services and what reaches the backend system**s.** This brings with it security implications that could impact services interacting with the application.

We have grouped all the security solutions and considerations described in this section into security domains and present their security requirements in Table1.

These security requirements can be implemented in different ways and often involves trade-offs. Decisions that involve trade-offs among constrained resources require tools, techniques and measurement to assist in decision making.

Table 1: SOA Security Domains and requirements

| Security Domains | Related SOA elements | Requirements |
|---|---|---|
| Message protection | -Transport<br>-Service communication protocol<br>-SOA external<br>-Applications front end's | -Confidentiality<br>-Integrity<br>-Authentication |
| Resource protection | - Service description<br>- Service registry<br>- Devices<br>- Applications front end | -Authorization<br>-Privacy<br>-Audit |
| Security properties Specification | -Service Registry<br>-Service Description<br>-Policy<br>-Business process | -SOA Security Policy, Business Security Policy<br>-Business process security specification |
| Security Negociation | -Service description<br>-Service registry<br>-Business process | -Negotiation Policy (Business contract, SLA)<br>-Security Properties dynamic discovery and binding |
| Security Management | -Management | -Security Monitoring,<br>-Security Governance<br>-Generating security implementations from abstracted security requirements |

## 2.4. SOA Maturity Models and Security

SOA is an evolutionary step-by-step approach to a new computing paradigm and infrastructure organization rather than a stand-alone software product. The evaluation of the level of development of SOA but also its implementation and usage, including Security features, can be supported by Maturity Models [22].

We have reviewed in [23] several SOA Maturity Models produced by Industry and academic contributors. Their common characteristic is that they define the process of SOA implementation using different SOA maturity levels. Each maturity level is an enhancement of the previous level and represents a set of criteria that have to be fulfilled during the process of SOA implementation. Those criteria or Key Indicators can be categorized in viewpoints: technical, organizational, governance, etc. We have noticed that none of these models presented security as a potential viewpoint. As a consequence, developing a Security viewpoint in an SOA Maturity Model and its associated metrics can support to evaluate and enhance the ability of SOA organizations to meet the objectives of security.

We represent in figure 2 a maturity model from the academic world [24] that is constituted of five maturity levels: Trial SOA, Integrative SOA, Administered SOA, cooperative SOA and on Demand SOA and five technical and organizational viewpoints: Service Architecture, Infrastructure, Enterprise Structure, Service Development and Governance.

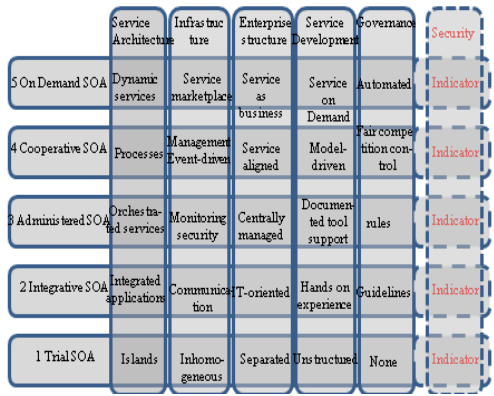

Fig. 2 Academic Maturity Model [24] extended by a security viewpoint.

Based on this model, we have represented security as a potential additional viewpoint. Its Key Indicators need to be determined to be coherent with other viewpoints indicators. Then Security metrics need to be defined in order to provide a yardstick through which organizations can: (i) compare with one another upon established baseline, (ii) auto evaluate themselves and implement security measures in a gradual approach.

## 3. Security metrics: what, why and how?

Security metrics are defined to provide a quantitative and objective basis for security assurance. They focus on aspects of the system that contribute to its security and involve the application of a measuring method for entities that have a measurable Security property. Moreover, they should enable an organization to assess how well it achieves its security objectives and they lead to actions to improve the security program of an organization [3].

We will explore in the next section two alternatives for defining security metrics: (i) Using metrics available in standards assessment frameworks, (ii) Deriving metrics from security goals using guiding and structured methods.

### 3.1. Standard Security Assessment Frameworks and metrics

Security Assessment frameworks help organizations assess their security risks, implement appropriate security controls, and comply with governance requirements as well as information security regulations [25]. According to different studies, these frameworks are based on metrics that can be categorized in organizational, technical or operational types (See table 2).

Table 2: Security Metrics Classification [4], [5], [6]

| Metric type | Description | Example | References |
|---|---|---|---|
| Organizational Metrics | They measure and monitor the effectiveness of programs and organizational processes | Compliance against International Standards, Process maturity statistics | *Systems Security Engineering Capability Maturity Model [26] *ISO 27002 [27] |
| Technical Metrics | They measure and compare technical objects | EAL ratings (Evaluation Assurance Level), Number of vulnerabilities | *Common Criteria [28] *Common Vulnerabilities and Exposures (CVE) list [29] |
| Operational Metrics | They measure and manage the risks of operating environments | Spam/viruses caught, security support tickets | NIST's Practices and Checklists Implementation Guide [30] |

These assessment frameworks suffer from the following: (i) they focus mainly on providing sets of controls, but the measurement of the quality and applicability of these controls is not handled in detail ; (ii) their metrics are associated with organizational security program; therefore its results can't be meaningfully comparable across different organizations ; (iii) Performance measurements of individual IS components like networks or software does not depict the overall security posture of an organization.

In order to define security metrics that can fit to the context of SOA organization where it is applied and overcome these limits, an approach for tailoring and refining standard security metrics can be helpful.

For that purpose we have selected two best practices standards that can support an organization to assess progress toward setting and meeting the stated security goals: ISO 27002 [27] and SSE-CMM [26]. ISO 27002 contains suggestions for security controls covering confidentiality, integrity and availability aspects from 11 areas of Security like Security Policy, Access Control, Compliance. The control's objective is to ensure that generally accepted good practices are met for each category in each field of information security. SSE-CMM is a tool to appraise and improve an organization's security engineering practices. It defines six maturity levels containing Generic Practices (GP) Grouped into logical areas called "Common Features". It defines also 22 Process Areas including Base Practices (BP). The SSE-CMM relates both base and generic practices and helps check that an organization is able to perform a particular security activity.

In [23], where we present a maturity model for SOA Security, we have justified in details the combined use of these security frameworks to assess security practices. Security measures from these frameworks need to be selected, tailored and mapped to SOA Security Domains and then refined into metrics based on a structured method.

### 3.2. Methods for deriving Security Metrics from security goals

After a literature review, we have selected and compared three methods that support metrics derivation from goals: GQM (Goal, Question, Metric) approach [31], GAM (Goal, Argument Metric) [32] and BSc (Balanced Scorecard Framework) [33].

GQM Approach provides a method for defining goals, refining them into questions and then defining metrics and finally data to be collected. GAM is a goal-oriented methodology for defining measurement plans. In GAM, the goals and sub-goals are represented as claims and then the analysis focuses on identifying which data and which properties of the data (further sub-goals) are needed to demonstrate these claims. BSc is a multidimensional framework for describing, implementing and managing strategy at all levels of an enterprise by linking objectives, initiatives and measures to an organization's strategy.

**Analysis:**
Considering the purpose and the general approach (top-down derivation and bottom-up interpretation) GQM and GAM look the same. The differences relate to the way of defining and maintaining the relationship between the measurement goals and the metrics. In GAM, the goals and sub-goals are represented as claims and then the analysis focuses on identifying which data and which properties of the data (further sub-goals) are needed to demonstrate these claims whereas in GQM referring to a goal, several questions are defined in such a way that obtaining the answers to the questions leads to the achievement of the measurement goal then based on the questions, metrics are defined, which provide quantitative information then treated as answers to the questions [32].

GQM Goals are referred to a project, while BSc goals are referred to a certain perspective and a certain particular tier in the organizational pyramid (hierarchy). Besides, GQM can be defined as a technique for deriving quantitative measures from a list of goals while BSc can be viewed as performance management framework that uses a GQM-like technique to derive the indicators [33].

GQM, GAM and BSc Structures similarities are presented in Table 3(adapted from [33]).

Table 3 : Comparison of GQM, GAM and BSc structures

| GQM | GAM | BSc |
|---|---|---|
| Goal | Claim | Goal |
| Question | Assertion | Driver |
| Metric | Metric | Indicator |

Both GAM and BSc are based on GQM Paradigm.

### 3.3. GQM Paradigm

GQM defines a top-down Measurement Model based on three levels [31]:

- **Conceptual level (GOAL)**

A goal is defined for an object for various reasons, with respect to various models of quality, from various points of view and relative to a particular environment. Object of Measurement can be: Products, Processes or Resources.

- **Operational level (QUESTION)**

A set of questions is used to characterize the way the assessment/achievement of a specific goal is going to be performed based on some characterizing model.

- **Quantitative level (METRIC)**

A set of metrics is associated with every question in order to answer it in a measurable way.

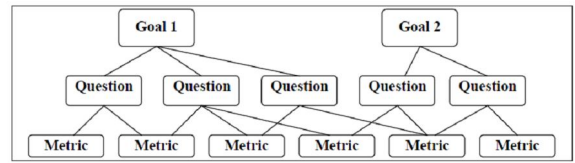

Fig. 3 GQM approach [31].

As the problem of defining 'good' questions is not easy, templates for defining more precisely the measurement goal have been introduced and are defined in a structured way, as presented in Table 4.

Table 4: Structure of GQM Template [31]

| Analyze | The object under measurement |
|---|---|
| For the Purpose of | Understanding, controlling, improving the object |
| With Respect To | The quality focus of the object that the measurement focuses on |
| From the viewpoint of | The people that measure the object |
| In The Context of | The environment in which measurement takes place |

## 4. Presenting the GQM approach proposed

### 4.1. Approach Model

Assessing the security of an SOA organization begins by defining relatively the security metrics appropriate to the context of its Information Systems. For that purpose, we propose a GQM approach to produce security metrics for an SOA organization based on its maturity level and on related security indicators. Our contribution's model is represented in figure4.

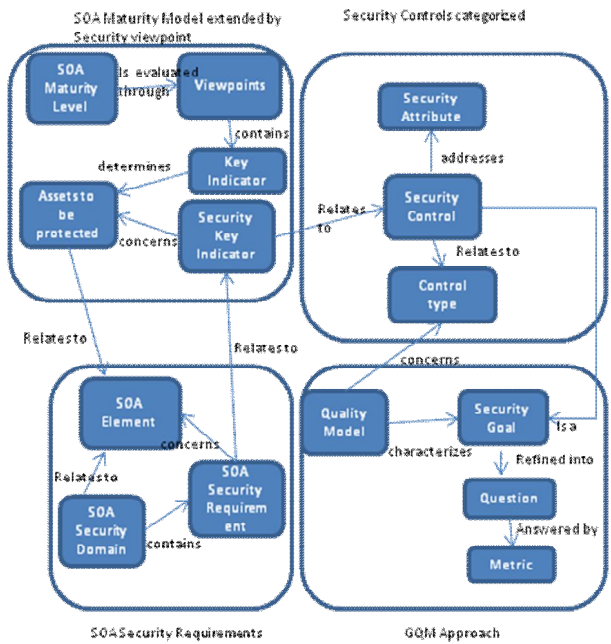

Fig. 4 Model of the approach proposed.

**Components of the approach :**

- SOA Maturity Model will help us precise the context of assessment.
- SOA Security domains and requirements from the analysis presented in 2.3 will support defining Key Indicators.
- Selected security controls (mainly from ISO 27002) mapped to SOA Security domains will help us define Security goals.
- Quality Model will characterize security goals using the implementation guidance from ISO 27002 and generic Practices from SSE-CMM Capability levels.

**Constraints :**

- We are not interested in security of Service Oriented Application components but only on service Artifacts and infrastructures. We have particularly relied on web services for defining SOA Security domains and requirements.
- We will focus on Access control features which rely on three main security attributes which are authentication, authorization and audit.
- SSE-CMM contains 12 common features to assess a security process capability. We will focus on the first and 2nd common features of SSE-CMM (Base Practices Performed and Planned Performance) for assessing process-like goals.

For each SOA maturity level, Security Key Indicators will be defined based on security context provided by Maturity Model viewpoints and SOA Security requirements. Key Indicators will be then related to selected security controls that will be considered as the goals that GQM method will refine into questions and metrics. These security controls address specific security attributes (authentication, authorization, integrity, etc) and relate to a control type (Policies, Processes, Software, hardware functions, etc).

In order to verify goal achievement, questions have to be derived which take the different aspects of a goal into account. For that purpose we will use GQM templates (Table 4) that support defining measurement goals in a structured way. Using the mapping between security controls types and GQM object of measurement presented in Table 5, we have produced the quality model, presented in Table 6, which will be used to characterize security goal.

Table 5: Mapping between Security control types and GQM object of Measurement

| Security Control type | GQM Measurement Object |
|---|---|
| Policy, Procedure | Product |
| Process | Process |
| Organizational structures, Software and hardware functions | Resource |

Table 6 : GQM Templates for Security controls

| Analyze | Security control process-like | Security control Resource-like | Security control Product-like |
|---|---|---|---|
| For the Purpose Of | Improving | Controlling | Controlling |
| With Respect To | Security capability | Effectiveness | Alignment Consistency |
| From the viewpoint of | Process owner | Security Administrator | Business owner |
| In The Context Of | SOA Maturity Level and SOA Security Domain | SOA Maturity Level and SOA Security Domain | SOA Maturity Level and SOA Security Domain |

For product or resource- like goals, related questions will be defined based on the implementation guidance provided by ISO 27002. For process-like goals, will be used capability levels and their generic practices provided by SSE-CMM. Based on the questions, metrics can be defined which are the basis for answering the questions. The questions in turn can be understood as guideline of how to interpret the measured metrics.

4.2. Approach Steps

Our contribution is a Top-Down approach for defining SOA security metrics from Security goals that are appropriate to an SOA context using GQM Method. It is divided into the following steps:

1. Step 1: To construct a Security viewpoint in the SOA Maturity Model by linking each maturity level to its related Security requirements grouped into security domains. The mapping is deducted from the context analysis of other viewpoints. The security requirements will be considered as KI (see table 7)
2. Step 2: To select security controls from ISO 27002 and literature that are related to Key Indicators defined in step 1(see table 8). These controls will be considered as the goals that GQM method will use to derive metrics.
3. Step 3: To refine Security goals into questions according to Control types and using a quality model defined based on templates provided by GQM method (Tables 5 and 6).
4. Step 4: To define security metrics. Based on the questions, metrics can be defined which are the basis for answering the questions.

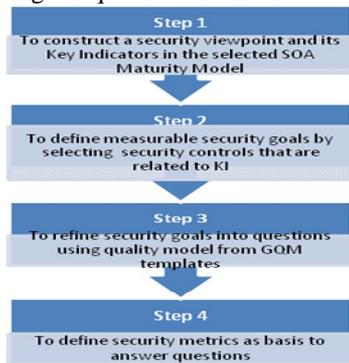

Fig. 5 Steps of the GQM approach proposed.

In the next section, we will use the approach to define security metrics related to access control features. For that purpose, we have selected an SOA maturity model which is isoamm [24].

## 5. Using the approach to define Security Metrics

5.1. Building a Security Viewpoint in the SOA Maturity Model

For each maturity Level, we need to define the SOA Security Context related to Access Control that is provided by other viewpoints: Service Architecture, Infrastructure, Governance, Service Development, Enterprise structure.

**Trial SOA:** SOA is in its infancy. Services start interacting in a point to point way. Security training and awareness related to SOA environment should start at this level. Basic security services can be setup at network and transport level through Firewalls and VPNs for instance. Devices access should also be controlled since this level.

**Integrative SOA:** This level is characterized by the integration of services through the Service Bus. Security controls that are applicable concern service communication protocol level and service resource access control. Security Policy should start to be formalized at this level.

**Administered SOA:** This level is characterized by service orchestration. Security properties need to be decoupled from application. This will support safe and correct processing of orchestrated services.

**Cooperative SOA:** This level is characterized by business processes, service interaction with application's front ends and by the use of SLAs to guarantee service quality. SOA environment should support the definition of security requirements at business process level and the use of SLAs. Besides, Security at this level must take into considerations front end's applications.

**On Demand SOA:** This level is characterized by its dynamicity and by SLA auto negotiation.

Following the $1^{st}$ step of the approach, we propose the following security viewpoint and its Key Indicators related to Access Control of each maturity level of the SOA Maturity Model.

Table 7: SOA Security viewpoint and its KI related to Access Control

| Maturity Level | SOA Security Domains | Security viewpoint related to Access Control |
|---|---|---|
| 5- On Demand SOA | Security negotiation | Security Properties dynamic discovery and binding |
| | Security Management | Monitoring of SLA Compliance to security rules |
| 4- Cooperative SOA | Resource protection | Applications front end AC |
| | Security properties specification | Security properties defined in SLA |
| | | Business process security specification language |
| | Security Management | Generating security implementations from abstracted security requirements |
| 3-Administered SOA | Security properties Specification | AC assertions in Service security Policy |
| | | AC properties in Service description and in registries |
| | Resource Protection | Service Security Policy AC |
| | Security Management | Service AC Monitoring |
| 2- Integrative SOA | Security Properties Specification | Access Control Policy Definition |
| | Message Protection | Message AC at Service Communication protocol Level |
| | Resource Protection | Service description AC |
| | | Registry AC |
| | Security Management | Infrastructure AC Monitoring |
| 1- Trial SOA | Message protection | Message AC at Transport level |
| | Resource protection | Devices Access Control |
| | SecurityManagement | Human Resource Security |

The next step is to define the goals related to Access Control KI and to refine them into questions and metrics.

## 5.2 Developing Goals

Based on ISO 27002 Security controls and other controls from the literature (marked with (*)), we have developed security goals, presented in Table 8, that we have related to SOA Access Control KI.

Table 8 : Goals related to SOA Access Control Key Indicators

| SOA AC Key Indicators | Security goals from ISO 27002 and literature(*) |
|---|---|
| AC Properties dynamic discovery and binding | G1- Use of Security Semantic Annotation in Service description and registry (*), G2- Architectural requirements for AC Properties Negotiation (protocol for negotiation, negotiation service, mediation service, auditing service)(*) |
| Monitoring of SLA Compliance to Security Rules | G1- Formal definition and expressiveness of SLA; G2- Use of Measurement and management infrastructure |
| Security properties defined in SLA | G1-Access Control Policy, G2- Policy on the use of cryptographic controls, G3- Security of net-work services, G4- Secure Business Information systems, G5- Secure Electronic commerce, G6- Secure On-Line Transactions, G7- Addressing security when dealing with customers, G8- Security rules definition for service selection at runtime (*) |
| Business process security specification | G1- Security requirements analysis and specification, G2- Languages to specify business process and its security constraints (*) |
| Generating security implementations from security requirements | G1- Use of Model Driven Security to automate the generation and re-generation of technical security enforcement from generic models (*) |
| Applications front end AC | G1- Identification of risks to related external parties, G2- privilege management, G3- User authentication for external connections, G4- Secure Log-on Procedures, G5- User identification and authentication, G6- Password management system |
| AC properties in Service description and registries | G1- Service delivery |
| Service AC Monitoring | G1- Monitoring and review of third party services, G2- Monitoring system use, G3- Key Management, G4- Control of technical vulnerabilities |
| AC assertions in Service security Policy | G1- Addressing security in third party agreements, G2- Security of network services |
| Access Control Policy Definition | G1- Classification guidelines, G2- Information labeling and handling, G3- Information exchange policies and procedures, G4- Policy on use of network services |
| Message AC at Service Communication protocol Level | G1- Secure Electronic messaging, G2- Network connection control, G3- Network routing control |
| Service description AC | G1- Information access restriction, G2- Access control to program source code |
| Registry AC | G1- User registration |
| Infrastructure AC Monitoring | G1- Audit Logging, G2- Clock synchronization |
| Message AC at Transport level | G1- Node authentication, G2- User identification and authentication |
| Devices AC | G1-segregation in networks and use of System utilities, G2- Remote diagnostic port protection |
| HR Security | G1-Information security awareness and training |

## 5.3. Refining Security Goals into questions and deriving metrics

For the sake of brevity, we present in Table 9 metrics related to following SOA Security KI: Access Control Policy Definition, Message Service Communication Protocol Access Control, Service Description Access Control, and Access Control Monitoring. Their related questions are detailed in appendix.

Table 9: Security Metrics for selected SOA Access Control KI

| SOA AC KI | Related Security Goals | Metrics |
|---|---|---|
| Access Control Policy Definition | Classification Guidelines | -% applications classified -application classification process quality -% critical applications |
| | Information labeling and handling | -% applications with protection plan procedures -application protection plan process quality |
| | Information exchange policies and procedures | -Access control procedure for exchanged information in security policy. -use of cryptographic techniques stipulated in security policy -security policy alignment with legal requirement |
| | Policy on use of network services | -% services allowed to be accessed covered by security policy -Service authorization procedures in security policy |
| Message AC at Service communication protocol | Secure Electronic messaging | - nb of message AC incidents - % of services with weak authentication techniques |
| | Network connection control | -% service whose access rights are aligned with security policy -use of network gateway to restrict service connection -% of message restricted by network gateway according to access control policy |
| | Network routing control | -use of routing controls for network -% routing control aligned with security policy |
| Service description Access Control | Information access restriction | -use of menu to control access to service functions -% of service functions modified without control access |
| | Access control to program source code | -change management procedures - % service managed according to the change management procedure |
| AC Monitoring | Audit Logging | -Use of Audit Logs -Detail of audit log |
| | Clock synchronization | -synchronize clock |

## 5.4. Detailing metrics

We will detail in this section metrics of the following Key Indicator: 'Message Access Control at Service communication protocol Level' which has three main security goals: Secure electronic messaging, Network connection control, Network routing control.

**Key Indicator**: Message Access Control at Service communication protocol Level

➢ **Goal 1: Secure electronic messaging**
- **Objective:** Information involved in electronic messaging should be appropriately protected
- **Quality Focus**: To control the effectiveness of the goal
- **Metrics:**
  *Metric1 : nb of message access control incidents*
  *Implementation evidence:*
  1- Are messages protected from unauthorized access with appropriate access control mechanisms? Answer : Yes or No
  2- Does the organization collect and review audit logs associated with unauthorized access to messages? Answer : Yes or no
  3- How many incidents related to unauthorized access to messages were logged within the reporting period? Answer : (number)
  *Target:* the measure should be as low as possible; target defined by the organization
  *Metric2: % of services with weak authentication technique*
  *Implementation evidence:*
  1- Are strong levels of authentication controlling access of messages from publicly accessible networks? Answer: Yes or no
  2- How many services are in the inventory? Answer : ( number)
  3- How many services use weak authentication techniques? Answer : (number)
  *Formula:* Number of services with weak authentication techniques/Number of services in the inventory *100
  *Target:* the measure should be low percentage defined by the organization

➢ **Goal 2: Network routing control**
- **Objective:** Routing controls should be implemented for networks to ensure that service connections and message flows do not breach the access control policy of the business application
- **Quality Focus**: To control the effectiveness of the goal and its alignment with security policy
- **Metrics:**
  *Metric1: use of routing controls for network*
  *Implementation evidence:*
  1- Do you implement routing controls for networks ? Answer : Yes or No
  *Metric2: % routing control aligned with security policy*
  *Implementation evidence:*
  1- Are these routing controls aligned with security policy? Answer : yes or no
  2- How many routing controls for networks are implemented? Answer : (number)
  3- How many routing controls are aligned with access control policy? Answer: (number)
  *Formula:* Number of routing controls aligned with access control policy/Number of routing controls implemented *100
  *Target:* the measure should be high percentage defined by the organization

➢ **Goal 3: Network connection control**
- **Objective:** To restrict the capability of services to connect to the network in line with the access control policy and requirements of the business applications.
- **Quality Focus**: To control the effectiveness of the goal and its alignment with security policy
- **Metrics:**
  *Metric1:* **% service whose access rights are aligned with security policy**
  *Implementation evidence:*
  1- Are network access rights of services maintained and updated according to the access control policy?
  2- How many services are in the inventory? Answer: (number)
  3- How many services have their access rights aligned with access control policy? Answer : (number)
  *Formula:* Number of services with access rights aligned /Number of services in the inventory *100
  *Target:* the measure should be high percentage defined by the organization.
  *Metric2:* **use of network gateway to restrict service connection.**
  *Implementation evidence:*
  Is there a network gateway that can restrict the connection capability of service? Answer : yes or no
  *Metric 3:* **% of message restricted by network gateway according to access control policy**
  *Implementation evidence:*
  1- How many messages are restricted by network gateway during a specified period? Answer : (Number)
  2- How many messages are restricted according to access control policy? Answer : (Number)
  *Formula:* Number of Messages restricted according to access control policy/Number of messages restricted *100
  *Target :* the measure should be high percentage defined by the organization

## 6. Discussion

We have presented a GQM approach to define security metrics to evaluate access control features of SOA Systems. Once defined, these security metrics can be used to evaluate SOA Security through the following assessment process.

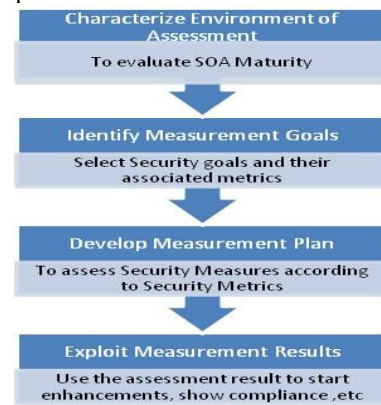

Fig. 6 Assessment Process using the security metrics.

Evaluating SOA security can be useful for organizations that use SOA as a basis for their collaborations in order to: (i) Use the assessment result to ensure the safety of their SOA collaborations; (ii) Enhance their security implementations to be aligned with their current or target SOA Maturity Level; (iii) Improve the quality of their security processes according to the security capability level they are targeting.

This work has advantages and limitations.

Let's start by the limitations. Practically, the use of SOA Maturity Model can have some limitations because if used in an inter organizational context, there is a need for consensus to use one SOA maturity model and especially the one that we have proposed. Besides, there is no certainty about the completeness of security metrics we have defined because one of the difficulties of GQM approach is to ensure that a quite detailed quality model is defined to characterize the achievement of goals.

The advantage of our work is that we had tried to mitigate these limitations. Concerning the first one, we have selected a maturity model that is based on CMM concepts and that use technical and organizational dimensions to evaluate the Maturity of SOA. Even if used with another SOA Maturity Model, there is possibility to define mapping relations between them but the reference will be the Maturity Model we have selected. Concerning the 2$^{nd}$ limitation, we have relied for the definition of our quality model on security standards that are well recognized in the field of security assessment and we have limited the scope of our work to access control characteristics and to SOA Artifacts and infrastructures.

## 7. Conclusion

In the new ecosystem proposed by SOA, multiple organizations work collaboratively creating a degree of programmatic complexity where security is a big concern. Ensuring the safety of collaborations is a challenge for organizations as they must have sufficient degree of Trust in other organizations to form a basis for willingness to engage in the interactions.

In this paper we have proposed a Goal Question Metric approach to derive security metrics from SOA Security Key Indicators (especially Access control Characteristics) and their related Security goals. Metrics that we have defined and their implementation evidence can be a support for organizations to assess their SOA Security especially for access control features, to ensure and to enhance the safety of collaborations based on SOA with other organizations.

## Appendix

**Questions derived for SOA Access Control Goals**

| KI | Goals / Quality Focus | Related questions |
|---|---|---|
| AC Policy Definition | **Goal:** Classification guidelines **Quality Focus:** To control and to improve the process | 1) Is the information classified in terms of its value, sensitivity, legal requirements and criticality to the organization? 2) Are there resources allocated to classify the Information? 3) Are responsibilities assigned to classify the Information? 4) Is the process of classifying information documented? 5) Are tools provided for classifying the Information? |
| | **Goal:** Information labeling and handling **Quality Focus:** To improve the process and to control the alignment of security control | 1) Is a set of Procedures for information labeling and handling developed? 2) Are there resources allocated to develop procedures for information labeling and handling? 3) Are responsibilities assigned to develop procedures for information labeling and handling? 4) Is the process of developing procedures for information labeling and handling documented? 5) Are tools provided for developing procedures for information labeling and handling? 6) Are procedures for information labeling and handling implemented in accordance with the classification scheme? |
| | **Goal:** Information exchange policies and procedures **Quality focus:** To control the consistency and alignment of control | 1) Does security policy stipulate procedures to protect exchanged information from unauthorized access? 2) Does the security policy stipulate to use cryptographic techniques to protect information? 3) Is security policy aligned with any relevant legal requirement? |
| | **Goal:** Policy on use of network services **Quality Focus:** To control the consistency and alignment of security control | 1) Does the security policy cover the networks and network services which are allowed to be accessed? 2) Does the security policy cover authorization procedures for determining who is allowed to access which networked services? 3) Does the security policy cover management controls and procedures to protect access to network connections and network services? 4) Is this security policy aligned with business policy? |

| KI | Goals/Quality Focus | Related questions |
|---|---|---|
| Message Service Communication Protocol Access Control | **Goal:** Secure Electronic messaging **Quality Focus:** To control the effectiveness of the security control | 1) Are messages protected from unauthorized access with appropriate access control mechanisms? 2) Does the organization collect and review audit logs associated with unauthorized access to messages? 3) How many incidents related to unauthorized access to messages were logged within the reporting period? 4) Are strong levels of authentication controlling access of messages from publicly accessible networks? 5) How many services are inventoried 6) How many services use weak authentication techniques? |
| | **Goal:** Network connection control **Quality Focus:** To control the effectiveness and alignment of the security control | 1) Are network access rights of services maintained and updated according to the access control policy? 2) How many services are in the inventory? 3) How many services have access rights aligned with access control policy? 4) Is there a network gateway that can restrict the connection capability of service? 5) How many messages are restricted by network gateway during a specified period? 6) How many messages are restricted according to access control policy? |
| | **Goal:** Network routing control **Quality Focus:** To control the effectiveness and alignment of security control | 1) Do you implement routing controls for networks? 2) Are these routing controls aligned with security policy? 3) How many routing controls for networks are implemented? 4) How many routing controls are aligned with access control policy? |

| KI | Goals / Quality Focus | Related questions |
|---|---|---|
| Service Description AC | **Goal:** Information access restriction **Quality Focus:** To control the effectiveness of security control | 1) Do you provide menus to control access to service functions and interfaces? 2) Do you control access rights of users and other services to service functions and interfaces? |
| | **Goal:** AC to program source code **Quality focus:** To control the effectiveness and alignment of security control | 1) Do you have a procedure for managing program source codes and libraries? 2) Are the program source codes and libraries managed according to the procedure? 3) Do you maintain audit log of all accesses to program source libraries? |

| KI | Goals / Quality Focus | Related questions |
|---|---|---|
| AC Monitoring | **Goal:** Audit Logging **Quality Focus:** To control the effectiveness of the security control | 1) Do you record audit logs? 2) Does your audit log contain user IDs? 3) Does your audit log contain dates, times, and details of key events? 4) Does your audit log contain records of successful and rejected system and date access attempts; 5) Does your audit log contain use of privileges |
| | **Goal:** Clock synchronization **Quality Focus:** To control the effectiveness of the control | Do you synchronize clocks of all relevant information processing systems within a security domain with an agreed accurate time source? |